\documentclass[a4paper,11pt]{article}
\pdfoutput=1
\usepackage{color}

\usepackage{jcappub}
\usepackage{amsthm}
\usepackage{graphicx}

\usepackage[normalem]{ulem}

\providecommand{\e}[1]{\ensuremath{\times 10^{#1}}}

\newcommand{\mpl}{{\ensuremath{M_\mathrm{Pl}}}}

\def\vev#1{ \left\langle #1 \right \rangle }

\renewcommand{\th}{{\ensuremath{^\mathrm{th}}}}

\newcommand{\ICsucc}{\mathrm{IC}_\mathrm{succ}}
\newcommand{\ICfail}{\mathrm{IC}_\mathrm{fail}}

\title{Inflating an Inhomogeneous Universe}

\author[1]{Richard Easther,}
\author[1]{Layne C. Price,}
\author[2]{and Javier Rasero}

\affiliation[1]{Department of Physics \\  University of Auckland \\ Private Bag 92019 \\  Auckland, New Zealand}
\affiliation[2]{Departament de F\'{i}sica Te\`orica and IFIC \\  Universitat de Val\`encia-CSIC \\ E-46100, Burjassot, Spain}

\emailAdd{r.easther@auckland.ac.nz}
\emailAdd{lpri691@aucklanduni.ac.nz}
\emailAdd{javier.rasero@uv.es}

\begin{document}

\abstract{While cosmological inflation can erase primordial inhomogeneities, it is possible that inflation may not begin in a significantly inhomogeneous universe.  This issue is particularly pressing in multifield scenarios, where even the homogeneous dynamics may depend sensitively on the initial configuration.  This paper presents an initial survey of the onset of inflation in multifield models, via qualitative lattice-based simulations that do not include local gravitational backreaction. Using hybrid inflation as a test model, our results suggest that small subhorizon inhomogeneities do play a key role in determining whether inflation begins in multifield scenarios.  Interestingly, some configurations which do not inflate in the homogeneous limit  ``succeed'' after inhomogeneity is included, while other initial configurations  which inflate in the homogeneous limit   ``fail'' when inhomogeneity is added.}

\maketitle
\flushbottom

\section{Introduction}
\label{sect:introduction}

The key predictions of inflation are strongly supported by observations of both large-scale structure \cite{Cole:2005sx,Eisenstein:2011sa,Giannantonio:2013uqa,Leistedt:2014zqa} and the cosmic microwave background, \emph{e.g.},  WMAP9 \cite{Bennett:2012fp,Hinshaw:2012fq}, ACT \cite{Sievers:2013wk}, SPT \cite{Story:2012wx,Hou:2012xq}, and \emph{Planck} \cite{Ade:2013xsa,Ade:2013rta}.  While inflation can ameliorate fine-tunings associated with the hot Big Bang, this presupposes that the inflationary mechanism  does not itself rely on fine-tuned fundamental parameters.\footnote{For a recent discussion of these problems see Refs~\cite{Ijjas:2013vea,Guth:2013sya,Linde:2014nna,Ijjas:2014nta}.} In particular, if inflation can only begin from a narrow range of possible configurations of the pre-inflationary universe, then the fine-tuning problems of the standard Big Bang have not been solved. 

Despite its importance, the inflationary initial conditions problem has received relatively little attention.  In some cases (\emph{e.g.}, new inflation \cite{Coleman:1973jx,Linde:1981mu,Albrecht:1982wi})  the initial conditions must be tuned in the purely homogeneous limit, but a complete treatment of the issue involves  the fully inhomogeneous Einstein equations.  This problem has been addressed for single-field inflation, with the conclusion that approximate homogeneity is needed over a volume of a few Hubble radii for chaotic inflation \cite{Goldwirth:1990pm,Goldwirth:1991rj}, whereas new inflation has an inflationary attractor \cite{Albrecht:1985yf,Kung:1989,Kung:1990,Feldman:1989hh}.  However, the inhomogeneous  dynamics of  multifield inflation are almost entirely unexplored.   Multifield inflation models have rich dynamics and are motivated from high-energy theory \cite{Grana:2005jc,Douglas:2006es,Denef:2007pq,Denef:2008wq}, making them relevant for studies of inflationary initial conditions.

We take the first step toward analysing this topic by performing lattice simulations of the pre-inflationary period for two-field hybrid inflation \cite{Kofman:1986wm,Linde:1993cn,Copeland:1994vg}.  This  is a qualitative exploration, as  our results are based on solutions of the  inhomogeneous scalar field dynamics in a homogeneous, expanding spacetime, ignoring both local gravitational backreaction and non-zero spatial curvature. 

By ignoring  inhomogeneities in the metric we are making the same assumptions that underpin a vast number of numerical studies of inflationary preheating.  We focus on initial configurations which represent small perturbations to a homogeneous background, as the small initial field gradients  ensure that our analysis is self-consistent. Futhermore, this approach allows us to take advantage of the mature  numerical tools that have been developed to analyze preheating \cite{Felder:2000hq,Frolov:2008hy,Easther:2010qz}.  Consequently, while this project represents a significant advance on previous studies of the initial conditions problem for multifield inflation, all of which have been performed in the purely homogeneous limit, it also paves the way for  analyses based on full numerical relativity.\footnote{Numerical solvers for the single field Einstein-Klein-Gordon equation in three dimensions are described by \cite{Goldwirth:1990pm,Goldwirth:1991rj,KurkiSuonio:1993fg}, and there has been recent progress in simulations of  highly inhomogenous bubble collisions for single field systems in one dimension \cite{Wainwright:2013lea,Wainwright:2014pta} but three dimensional, multifield scenarios with significant inhomogeneity are beyond the scope of currently available numerical tools.}

The chaotic nature of the homogeneous limit of multifield inflation \cite{Easther:1997hm,Easther:2013bga} means adjacent trajectories in phase space are highly divergent, but as field gradients contribute to the energy density,  field values at nearby {\em spatial} points cannot diverge by arbitrary amounts. This effect potentially ``focusses'' trajectories relative to the homogeneous limit, and our  lattice-based simulations let us explore the role of the gradient energy in the inhomogeneous evolution of this system. Using Monte Carlo explorations of the initial conditions space, we confirm that the qualitative consequences of the chaotic dynamics,  especially phase-space mixing, persist when moderate-to-large inhomogeneity is added. However, for  many initial configurations the inflationary outcome is not changed by  the addition of small amplitude inhomogeneities, demonstrating the focussing effect. Moreover, while many initial conditions that ``succeed'' in the homogeneous limit  do ``fail'' when inhomogeneity is included, we also see initial configurations which ``fail'' in the homogeneous limit that successfully  inflate when inhomogeneity is included.

\section{Model}
\label{sect:model}

We consider hybrid inflation \cite{Linde:1993cn,Copeland:1994vg} with two inhomogeneous scalar fields $\phi$ and $\psi$, in a flat FLRW universe with equations of motion
\begin{equation}
  \ddot \phi_i + 3H \dot \phi_i - \frac{1}{a^2} \nabla^2 \phi_i + \frac{\partial V}{\partial \phi_i} = 0 \, ,
  \label{eqn:kgeqn}
\end{equation}
where $a$ is the scale factor, $H=\dot a/a$ is the Hubble parameter, and a  subscript $i$ denotes the components of the vector $\{\phi,\psi\}$.  The inhomogeneous energy density is
\begin{equation}
  \rho(t,\mathbf x)=\frac{1}{2} \sum_i \left[ \dot \phi_i^2 + \frac{  (\nabla \phi_i)^2}{a^2} \right]+ V(\phi_i),
  \label{eqn:XXX}
\end{equation}
the pressure is
\begin{equation}
  p(t, \mathbf x) = \frac{1}{2} \sum_i \left[ \dot \phi_i^2 - \frac{1}{3} \frac{  (\nabla \phi_i)^2}{a^2} \right] - V(\phi_i),
  \label{eqn:XXX}
\end{equation}
and $a$ evolves according to the Einstein equations.  In general, the metric is spatially dependent, but  we set $a = \vev{a(t,\mathbf x)}$ and $H^2 =  \vev{\rho(t,\mathbf x)}/3$, where $\vev{.}$ indicates an integrated spatial average, which is an accurate approximation when the fields' variance is small.  Relaxing this assumption would require a full numerical treatment of the Einstein equations.

The generic hybrid inflation potential $V$ has the form
\begin{equation}
  V(\phi, \psi) = \Lambda^4 \left[ \left( 1 - \frac{\psi^2}{M^2} \right)^2 + U(\phi) + \frac{\phi^2 \psi^2}{\nu^4} \right] ,
  \label{eqn:hybridv}
\end{equation}
where $U(\phi)$ drives a sustained period of slow-roll inflation in the ``inflationary valley'' at $\psi \approx 0$, but is otherwise subdominant.  A graceful exit from inflation occurs when the effective mass of the waterfall field $m_{\mathrm{eff},\psi}^2 \equiv \Lambda^4 ( \phi^2/\nu^4 - 2/M^2)$ becomes imaginary.

We work with
\begin{equation}
  U(\phi)=\frac{\phi^2}{\mu^2},
  \label{eqn:uphi}
\end{equation}
although the \emph{Planck} results~\cite{Ade:2013rta} rule out slow-roll inflation with this form of $U(\phi)$, since it predicts $n_s>1$.  
However, the exact form of $U(\phi)$ is likely to have little impact on the initial conditions problem since, by hypothesis, it only dominates the potential in a small portion of field space and the multifield dynamics for this scenario are well-studied \cite{Lazarides:1996rk, Lazarides:1997vv, Tetradis:1997kp, Mendes:2000sq, Ramos:2001zw,Clesse:2008pf,Clesse:2009ur,Easther:2013bga}.  We use hybrid inflation as a toy model to illustrate the interesting multifield dynamics resulting from the interaction between the dynamical fixed points at $\psi=\left\{ 0, \pm M \right\}$ and the tachyonic instability points $\phi_\mathrm{crit}=\pm \sqrt{2} \, \nu^2/M$.
We note that a red scalar spectrum can be achieved through other choices of $U(\phi)$ \cite{Dvali:1994ms,Buchmuller:2014epa}, as does inflation during the waterfall transition \cite{Clesse:2008pf,Clesse:2010iz,Kodama:2011vs,Clesse:2013jra}.
\begin{table}
  \centering
  \begin{tabular}{ c |  c c c c  }
    \hline
    \hline
    Parameter  & $\Lambda$ & $M$ & $\mu$ & $\nu$ \\
    \hline
    Value $[\mpl]$& 6.8\e{-6} & 0.03 & 500.0 & 0.015 \\
    \hline
    \hline
  \end{tabular}
  \caption{Parameter values for the potential in Eqs~\eqref{eqn:hybridv}~and~\eqref{eqn:uphi}.  The overall energy scale is set by $\Lambda$, but the background dynamics are not affected by this choice.}
\label{table:params}
\end{table}

The numerical parameters  used in our simulations are listed in Table~\ref{table:params}.  The background dynamics are independent of the value of $\Lambda$, but we choose $\Lambda = 6.8\e{-6} \mpl$ to match the measured   amplitude of the scalar power spectrum \cite{Ade:2013rta}.  The onset of inflation in this model has been thoroughly investigated in the  homogeneous limit \cite{Lazarides:1996rk, Lazarides:1997vv, Tetradis:1997kp, Mendes:2000sq, Ramos:2001zw,Clesse:2008pf,Clesse:2009ur,Easther:2013bga} and surveys of initial conditions that consider both the initial velocities and field values \cite{Clesse:2008pf,Clesse:2009ur,Easther:2013bga} find that  hybrid inflation scenario  begins for a significant fraction of initial configurations.  However, the chaotic nature of the underlying dynamics ensures that the set of successful initial conditions is  fractal.

\section{Numerical Methods}
\label{sect:method}

\paragraph{Equations:} We solve the equations of motion \eqref{eqn:kgeqn} using  \textsc{LatticeEasy}\footnote{\url{www.cita.utoronto.ca/~felder/latticeeasy/}} \cite{Felder:2000hq} and assuming periodic boundary conditions $\phi_i(t,\mathbf x) = \phi_i(t,\mathbf x-\mathbf L)$, where $\mathbf L$ is the length of the spatial box defined by the simulation.  On a lattice, the Fourier transforms can be expressed as a finite series
\begin{equation}
  \phi_i (t, \mathbf x) = \int \frac{d^3 k}{(2 \pi)^3} \hat \phi_{i,\mathbf k}(t) e^{ 2 \pi i \mathbf k \cdot \mathbf x } \to \frac{1}{L^3} \sum_\mathbf{k} \hat \phi_{i,\mathbf k}(t)e^{ 2 \pi i \mathbf k \cdot \mathbf x } \, .
  \label{eqn:XXX}
\end{equation}
We specify  initial conditions in Fourier space, through $\hat \phi_{i,\mathbf k}(0)$ and its time derivatives $\partial_t \hat \phi_{i,\mathbf k}(0)$. In order to sample a large number of initial configurations, we restrict the inhomogeneity to one spatial direction, effectively assuming a translational symmetry in the orthogonal directions. We allow a small initial velocity in the $0\th$ mode, but set $\partial_t \hat \phi_{i,\mathbf k > 0}=0$.

\paragraph{Inhomogeneous $T_{\mu \nu}$:} With a single excited  mode, the initial field configuration is
\begin{equation}
  \phi_i(0,x) = \bar \phi_{i,0}\left[1 + A \sin \left( \frac{2 \pi n x}{L} \right) \right] \, .
  \label{eqn:initinhomog}
\end{equation}
Periodic boundary conditions require the lattice-size to be a  multiple of the wavelength $k=n/L$ for some integer $n$.  Without loss of generality, we can set the boxsize $L$  to the wavelength of the largest mode we excite, given our choice of periodic boundary conditions, and define $n_\mathrm{largest}\equiv1$.

We further assume that $A \lesssim 1$ to keep the backreaction on the scale factor initially small and maintain the self-consistency of Eq.~\eqref{eqn:kgeqn}.  The average initial energy density is
\begin{equation}
  \langle \rho_0 \rangle = \frac{1}{2} \sum_i \dot \phi_i^2 + \vev{ V(\phi_i)} + \sum_i \left( \frac{n \pi A \bar \phi_{i,0}}{a_0 L} \right)^2
  \label{eqn:XXX}
\end{equation}
and the initial average gradient energy density is suppressed by $(n A/L)^2$.  Likewise, the relevant contribution of the inhomogeneity to the trace and traceless parts of the stress tensor are proportional to the square of the field gradient and are also suppressed by this factor at the start of the simulation.  The momentum density is
\begin{equation}
  \mathcal{J}_j = -\sum_i \dot \phi_i \partial_j \phi_i,
  \label{eqn:XXX}
\end{equation}
which we keep small initially by considering only small initial velocities in the $0\th$ mode, combined with a small value for $nA/L$.  However, the spatial average of $\mathcal{J}_j$ over the simulation volume is strictly zero when the fields have the form in Eq.~\eqref{eqn:initinhomog}.

As the simulation progresses we do not require that the components of the energy-momentum tensor to remain small, but only require that $a$ and $H$ are well approximated by their spatial average.  To ensure that the inhomogeneous contributions to the stress-energy tensor do not induce a large backreaction on the subsequent evolution, we  require that each field's variance
\begin{equation}
  \mathrm{Var}_i \equiv \vev{\phi_i^2}-\vev{\phi_i}^2
  \label{eqn:XXX}
\end{equation}
 remains small throughout the simulation, since large values would require a more sophisticated analysis involving  solutions of the full Einstein field equations.  Solutions for which the variance exceeds $\mathrm{Var}_i \gtrsim 10^{-2} \; \mpl^2$ are dropped from our analyses, but in practice, almost none of the configurations we consider generate variances that cross this threshold, as the overall simulation time is relatively short. 

\paragraph{Initial inhomogeneity:} We  parametrize the initial inhomogeneity associated with a mode of wavenumber $k$ using $f$, the ratio of its wavelength $\lambda$ to the initial Hubble radius in the homogeneous limit:
\begin{equation}
  f  \equiv \frac{a_0 \bar H_0}{ k} ,
  \label{eqn:frac}
\end{equation}
where $a_0\equiv1$ is the initial scale factor and $\bar H_0$ is the initial Hubble parameter in the homogeneous limit, \emph{i.e.}, using only $\bar \phi_{i,0}$ and $\partial_t{\bar{\phi}}_{i,0}$. The initial horizon size in the homogeneous limit is related to the size of the simulation box by $2 \pi f = a_0 \bar H_0 L$.  We  consider only subhorizon perturbations with $f \lesssim 1$ since superhorizon inhomogeneities can be well modeled by a collection of homogeneous universes with different initial conditions, using the separate universe assumption.  Subhorizon inhomogeneities contribute only perturbatively to the Newtonian potential, which further justifies ignoring metric inhomogeneities.  Assuming we start in an almost-FLRW universe, these small-scale inhomogeneities can contribute only an effective pressure term with $w=1/3$ that cannot contribute to exponential expansion \cite{Green:2010qy}.

\paragraph{Ending condition:} We require $N_e \geq 55$ $e$-folds of accelerated expansion with $\epsilon < 1$, although this  limit is somewhat arbitrary as almost any specific initial condition yields either $N_e \gg 55$ or  $N_e \ll 5$.  We  follow previous analyses \cite{Clesse:2008pf,Clesse:2009ur,Easther:2013bga} by defining an initial configuration as ``successful'' if it gives $N_e > 55$, and as a ``failure'' if it does not.  In practice, almost all of the $e$-folds occur in the inflationary valley when $\psi\approx 0$, so we halt our simulations once $N_e > 5$ and $\langle \psi \rangle \approx 0$, \emph{i.e.}, once the trajectory has settled into the inflationary valley.  Alternatively, we stop the integration if $\rho < V(0,\phi_c) = 2 \Lambda^4 \nu^4 / M^2 \mu^2$, since the trajectory cannot enter the inflationary valley; these trajectories are by definition unsuccessful.  In Section~\ref{sect:results} we compare our results for initial configurations with non-zero gradient energies to the homogeneous cases studied in Refs~\cite{Lazarides:1996rk,Tetradis:1997kp,Lazarides:1997vv,Mendes:2000sq,Clesse:2008pf,Clesse:2009ur,Easther:2013bga}, which are in the ``homogeneous limit'' with $\nabla \phi_{i,0} \to 0$.

\section{Results}
\label{sect:results}

\subsection{Single excited mode}
\label{ssect:singlemode}

We begin by examining two specific background initial conditions, $\ICsucc$ and $\ICfail$, which have different homogeneous dynamics.  These correspond to choices of initial conditions for which $N_e > 55$ and $N_e <5$, respectively.  The specific choices are
\begin{eqnarray}
  \label{eqn:icsucc}
\ICsucc \equiv
\left[
  \begin{array}{c}
  \bar \psi_{0,\mathrm{succ}} \\
  \bar \phi_{0,\mathrm{succ}} \\
  \dot{ \bar \psi}_{0,\mathrm{succ}} \\
  \dot{ \bar \phi}_{0,\mathrm{succ}}
  \end{array}
\right]
  &=&
\left[
  \begin{array}{c}
  9.6405\e{-3} \; \mpl \\
  2.7359\e{-2} \; \mpl \\
  -1.0211\e{-10} \; M_\mathrm{Pl}^2 \\
  1.6059\e{-11} \; M_\mathrm{Pl}^2
  \end{array}
\right], \\
\ICfail \equiv
\left[
  \begin{array}{c}
  \bar \psi_{0,\mathrm{fail}} \\
  \bar \phi_{0,\mathrm{fail}} \\
  \dot{ \bar \psi}_{0,\mathrm{fail}} \\
  \dot{ \bar \phi}_{0,\mathrm{fail}}
  \end{array}
\right]
  &=&
\left[
  \begin{array}{c}
  1.0361\e{-2} \; \mpl \\
  2.7497\e{-2} \; \mpl \\
  -6.6330\e{-11} \; M_\mathrm{Pl}^2 \\
  7.3672\e{-11} \; M_\mathrm{Pl}^2
  \end{array}
\right].
  \label{eqn:icfail}
\end{eqnarray}
The combined initial kinetic and potential energy is $E_0 = 10^{-5} \, \mpl$ and quantum fluctuations at this energy should have minimal impact on the end-state of the background evolution \cite{Easther:2013bga}.  By adding initial inhomogeneity to these background field values, we find that the amount of inflation given by both of these trajectories can change drastically.

\begin{figure}
  \centering
  \includegraphics{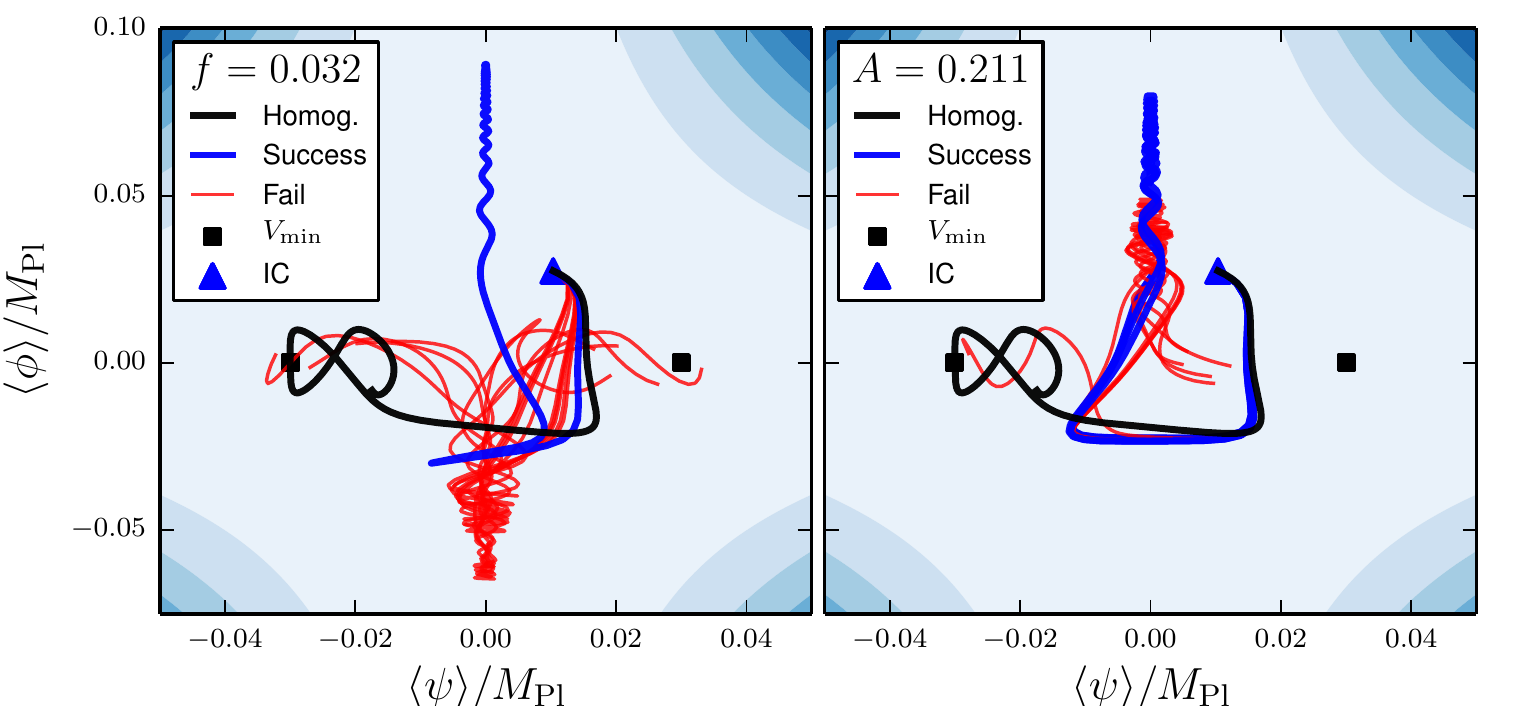}
  \caption{Spatially averaged solutions to Eq.~\eqref{eqn:kgeqn} for ten initial conditions with the same background field values $\ICfail$, with a sinusoidal inhomogeneity, of amplitude $A$ and comoving wavelength $\lambda/ 2 \pi = f/a_0\bar  H_0$, added in-phase to both fields.  In the homogeneous limit the background field values give $N_e \ll 55$.  The blue contours show the potential energy density $V$.  (\emph{Left}) The initial fraction $f$ is fixed and the amplitude is varied between $10^{-0.5} < A < 1$.  (\emph{Right})  The initial amplitude is fixed to $A= 1$ and the fraction $f$ is varied between $10^{-2} < f < 10^{-1}$.
  }
  \label{fig:failtrajs}
\end{figure}
\begin{figure}
  \centering
  \includegraphics{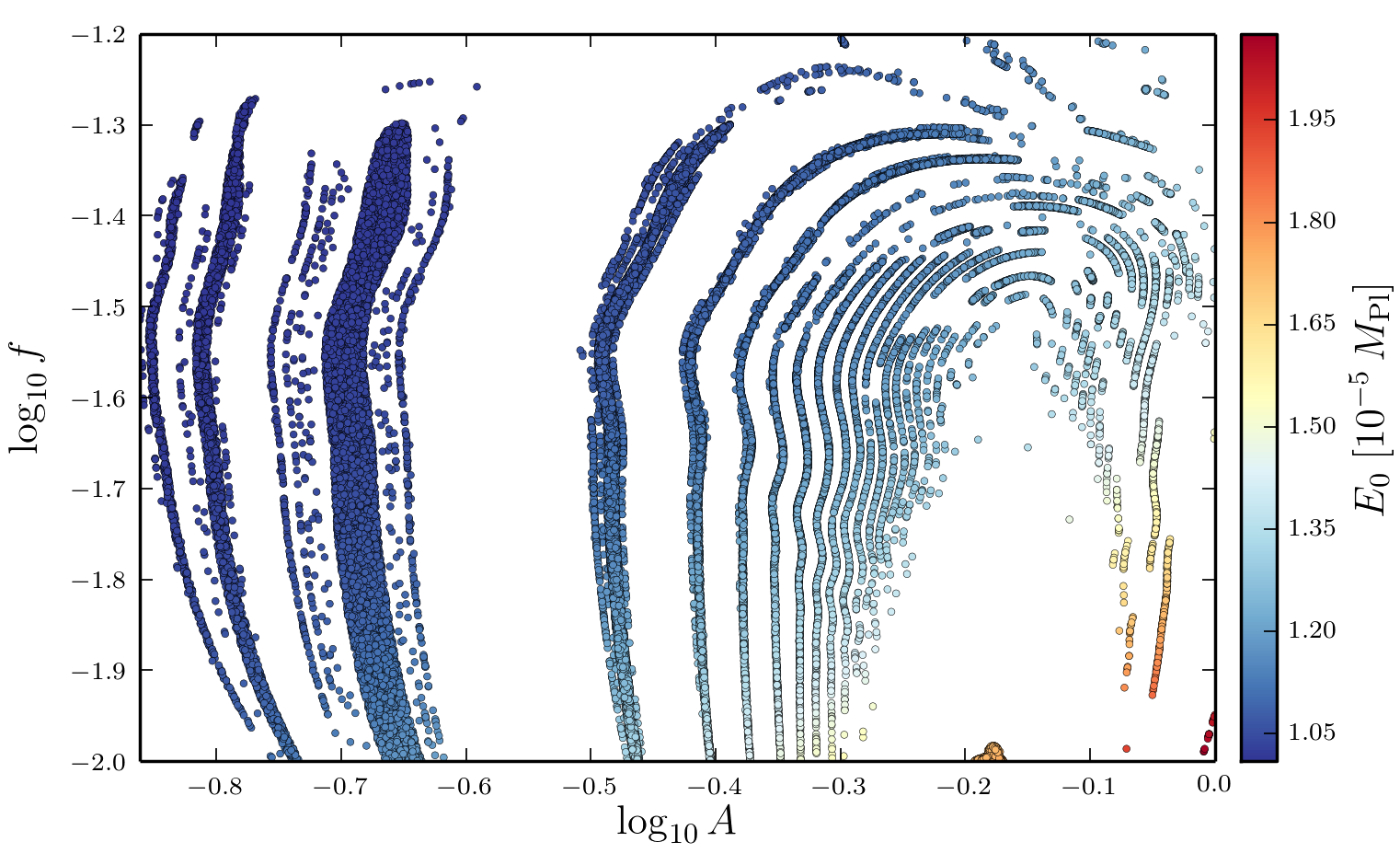}
  \caption{Distribution of independent initial conditions that succeed at generating $N_e>55$ $e$-folds of inflation, as a function of initial inhomogeneity.  The background field values $\ICfail$ are set so that in the homogeneous limit $N_e \ll 55$.  A sinusoidal inhomogeneity, with amplitude $A$ and comoving wavelength $\lambda/ 2 \pi = f/a_0\bar  H_0$, has been added, in-phase, to both fields.  The initial energy $E_0$ is indicated by color.}
  \label{fig:failepsvsfrac}
\end{figure}
\begin{figure}
  \centering
  \includegraphics{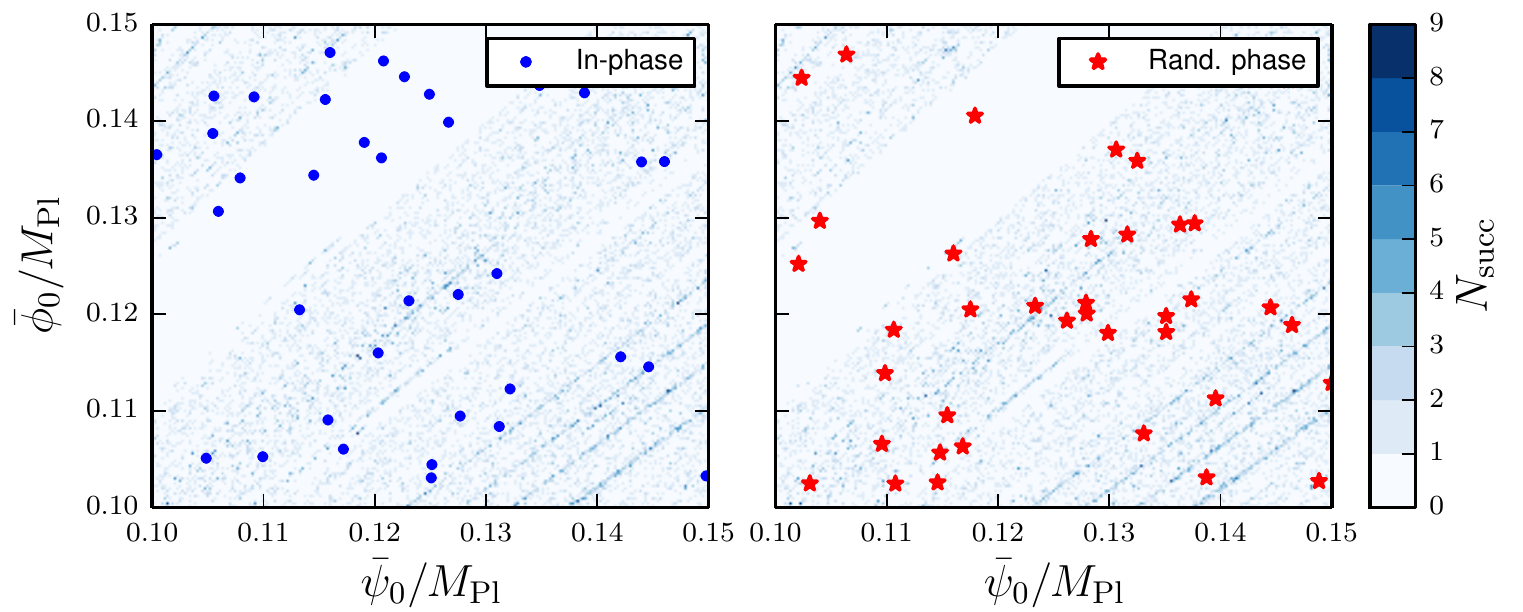}
  \caption{Addition of sub-horizon inhomogeneity can turn failure ICs into successful ICs.  We add one sinusoidal inhomogeneity (\emph{left}) in-phase and (\emph{right}) with a random phase difference to 50 randomly selected background ICs that fail in the homogeneous limit.  We find 38/50 and 39/50 change to ``successful,'' respectively, with no noticeable correlations between them.  (\emph{Background}) Histogram of successfully inflating ICs ($N_e \gg 55$) in the homogeneous limit, with the number of successful points per bin $N_\mathrm{succ}$ indicated by color.}
  \label{fig:hist_fail}
\end{figure}

In Fig.~\ref{fig:failtrajs} we plot the spatially averaged values of $(\phi, \psi)$ for a set of 10 configurations with background field values of $\ICfail$.  We perturb each configuration by adding a sinusoidal inhomogeneity with equal phases (as in Eq.~\eqref{eqn:initinhomog}) to both fields.  All of the plotted solutions explore much of the field space and most of the sampled trajectories are eventually captured in the inflationary valley at $\vev \psi =0$; however, only some trajectories stay there and inflate sufficiently.  Interestingly, this demonstrates that subhorizon inhomogeneity can actually cause inflation in scenarios that fail to inflate in the homogeneous limit.    Hybrid inflation has been shown to be chaotic, first in the $H \to 0$ limit in Ref.~\cite{Easther:1997hm}, then in the homogeneous limit in Refs~\cite{Clesse:2008pf,Clesse:2009ur,Easther:2013bga}.  Since phase-space mixing is a characterisitic of chaos, this is the first indication that this behaviour extends to the inhomogeneous Klein-Gordon equation.  The non-linear dynamics of multifield inflation may therefore have a significant effect on whether inflation successfully begins with from an inhomogeneous universe.

Since we are solving (1+1)-dimensional PDEs, the computational cost of evaluating each configuration is not excessive and we are able to generate large samples to test whether this behavior is generic.  Figure~\ref{fig:failepsvsfrac} shows $\mathcal O(10^6)$ Monte Carlo samples with background field values of $\ICfail$.  We again add a sinusoidal inhomogeneity with logarithmic priors on the parameters, $-2.0 < \log_{10} f < -1.2 $ and $-1.0 < \log_{10} A < 0.0$.  Trajectories with lower initial gradient energy than this do not deviate significantly from the homogeneous solution and fail.  The set of successful points has a fractal structure, similar to that seen with homogeneous hybrid inflation \cite{Clesse:2009ur,Easther:2013bga}.  Using the box-counting method \cite{theiler} we are able to determine a fractal dimension of $d=1.27$, $d=1.85$, and $d=1.25$, for the set of points in Fig.~\ref{fig:failepsvsfrac} that are successful, unsuccessful, and the boundary between the two, respectively.  This is convincing evidence that the dynamics remain chaotic in some regions of parameter space.

We also check that this behavior does not depend on the specific choice of $\ICfail$, by looking at scenarios with different background field values, and with the perturbations in $\phi$ and $\psi$ either in-phase or with an arbitrarily chosen phase difference.  The results of this investigation are presented in Fig.~\ref{fig:hist_fail}. We set the initial field velocities to zero and draw the background initial conditions from  $0.10 \, \mpl < \bar{\phi}_0, \bar{\psi}_0 < 0.15 \, \mpl$.  We then chose 50 configurations that fail in the homogeneous limit and add sub-horizon inhomogeneity with  arbitrary or zero phase differences between the modes in each field.   In most cases we find a mixture of inflationary and non-inflationary solutions at each point, demonstrating the generality of this phenomenon.

\begin{figure}
  \centering
  \includegraphics{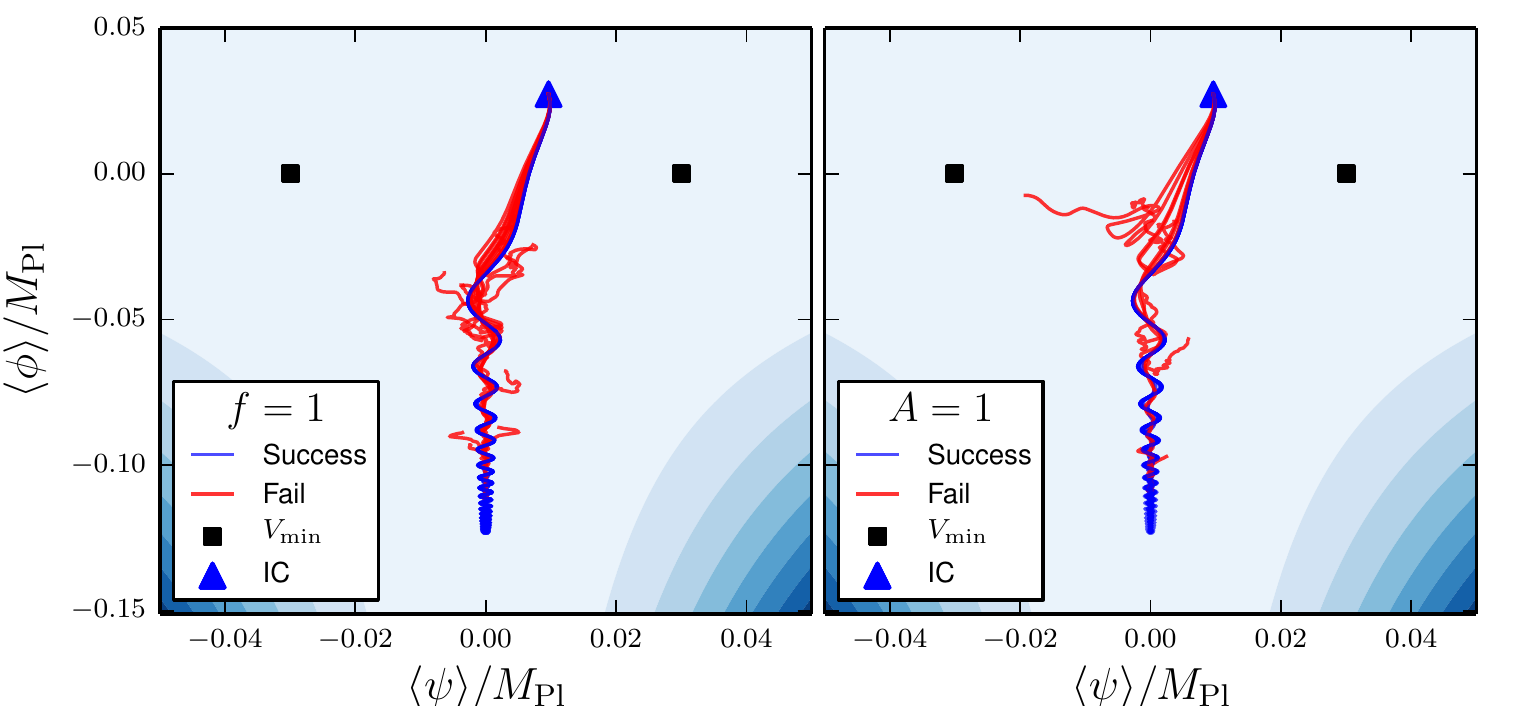}
  \caption{Spatially averaged solutions to the inhomogeneous Klein-Gordon equation for twenty initial conditions $\ICsucc$, and varying initial gradient energy, provided by a sinusoidal inhomogeneity, with amplitude $A$ and comoving wavelength $\lambda/ 2 \pi = f/a_0 \bar H_0$, in each field and with identical phase. The blue contours show the potential energy density $V$.  (\emph{Left}) The initial inhomogeneous wavelength is fixed to the initial Hubble scale $f=1$ and the amplitude is varied between $10^{-3} < A < 1$.  (\emph{Right})  The initial amplitude is fixed $A= 1$ and the wavelength is varied between $10^{-2} < f < 1$.
  }
  \label{fig:trajs}
\end{figure}

Conversely, Fig.~\ref{fig:trajs} displays the spatially averaged trajectories for initial configurations with background field values of $\ICsucc$ with the in-phase sinusoidal perturbations.
The spatially averaged field trajectories begin by following the homogeneous trajectory and oscillating around the inflationary valley at $\vev{\phi}<-\phi_\mathrm{crit}$.  For those configurations with initially small gradient energies,  accelerated expansion exponentially dampens the inhomogeneity, the trajectory is captured in the inflationary valley, and successfully inflates.  The successfully inflating trajectories in Fig.~\ref{fig:trajs} are nearly indistinguishable from each other and from the field-space trajectory of the homogeneous solution. With larger gradient energies --- obtained either by reducing the wavelength of the perturbation relative to the horizon or increasing the amplitude $A$ --- the inhomogeneity can pull the spatially averaged trajectory out of the valley.  These trajectories then evolve more-or-less directly to the minimum of the potential and will give at most a few $e$-folds during any transient inflationary phases.

\begin{figure}
  \centering
  \includegraphics{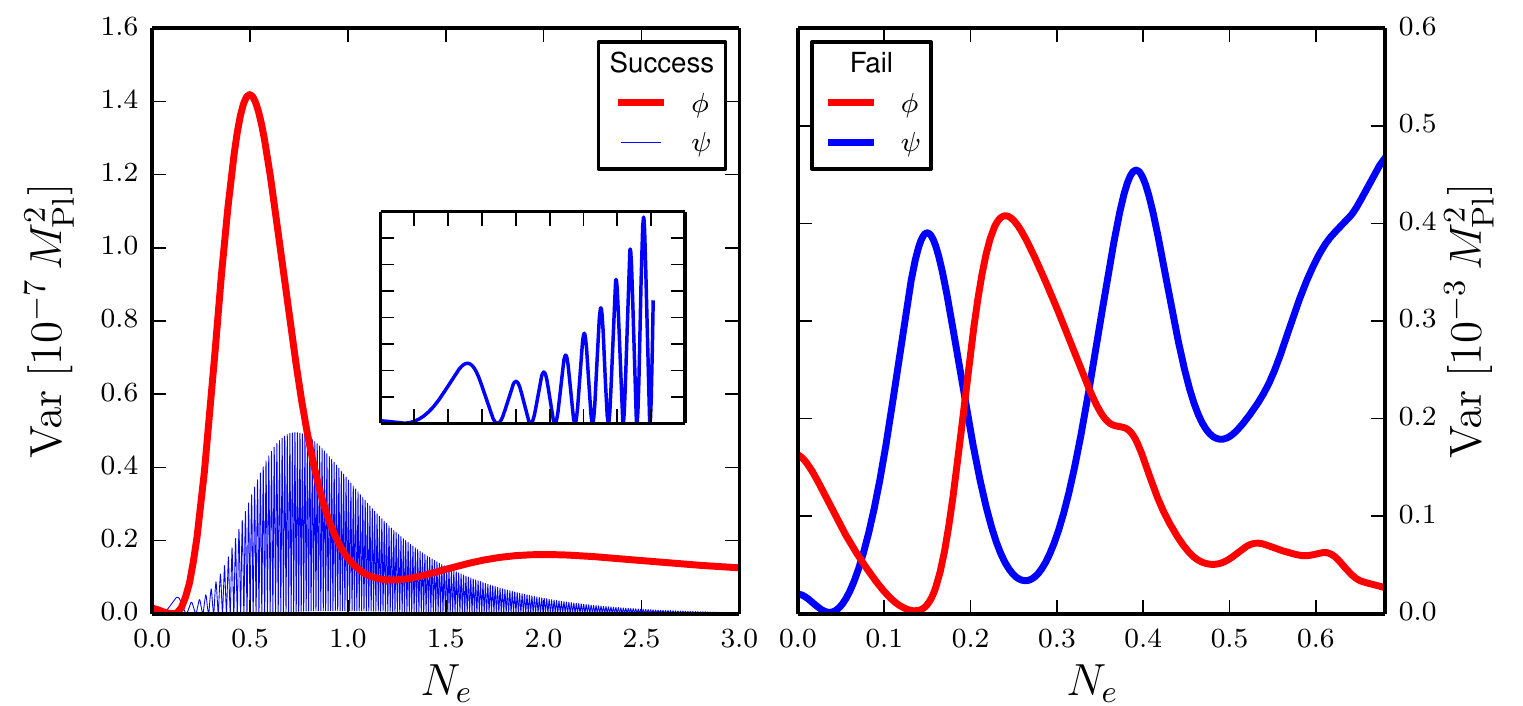}
  \caption{Variance in the fields $\mathrm{Var}_i = \vev{\phi_i^2}-\vev{\phi_i}^2$ as a function of $e$-folding $N_e$ for initial configurations with background field values of $\ICsucc$, as in Fig.~\ref{fig:trajs}.  (\emph{Left}) Configuration yielding $N_e>55$ with $f=1.890\e{-1}$ and $A=2.035\e{-3}$; (\emph{left inset}) zoom-in on $\psi$ for $N_e<0.403$. (\emph{Right}) Configuration yielding $N_e \ll 55$ with $f=3.556\e{-1}$ and $A=6.598\e{-1}$.
  }
  \label{fig:var}
\end{figure}

Fig~\ref{fig:var}  gives the variance $\mathrm{Var}_i$ for representative successful and unsuccessful solutions with initial background field values of $\ICsucc$.  For the successful case, the trajectory is captured by the false vacuum and oscillates around $\vev{\psi}=0$ with a frequency of $40$ [oscillations/$e$-fold].  Since $\mathrm{Var}_\psi \sim \vev{\psi}^2$ it therefore oscillates at a frequency of $80$ [oscillations/$e$-fold].  The variance in $\phi$ peaks only once at $N_e = 0.5$ and remains below $\mathrm{Var}_\phi < 10^{-7} \; M_\mathrm{Pl}^2$.  In the failing universe there is no extended period of oscillation around the false-vacuum at $\vev{\psi}=0$, so the oscillations in $\mathrm{Var}_\psi$ have a much smaller frequency.  After $N_e \gtrsim 0.5$ the inhomogeneities in $\psi$ start to grow substantially,  destablising the dynamics, and causing failure.

\begin{figure}
  \centering
  \includegraphics{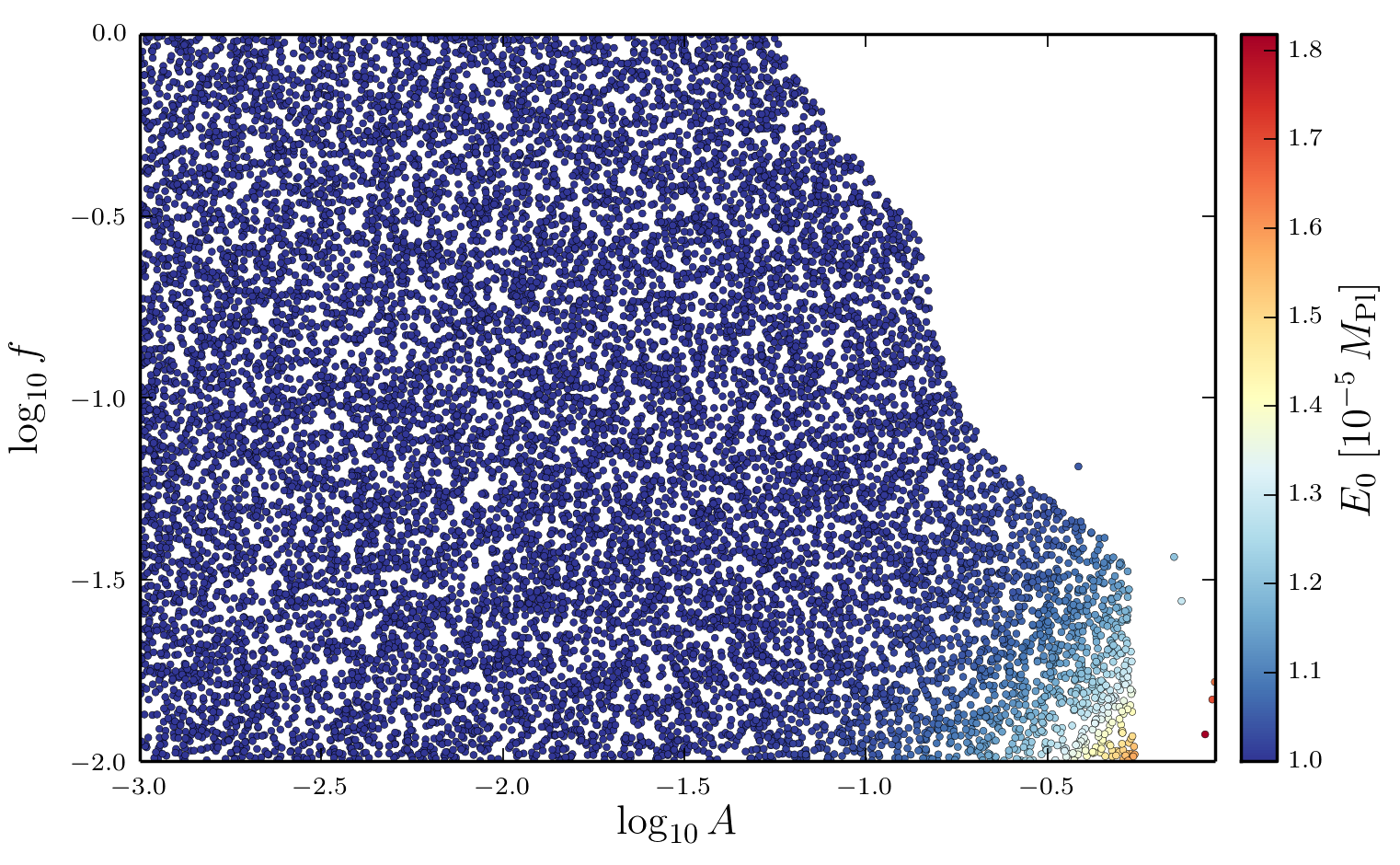}
  \caption{Distribution of independent ICs that succeed at generating $N_e>55$ $e$-folds of inflation, as a function of initial inhomogeneity.   The background field values $\ICsucc$ are set so that in the homogeneous limit $N_e>55$.  A sinusoidal inhomogeneity, with amplitude $A$ and comoving wavelength $\lambda/ 2 \pi = f/a_0\bar  H_0$, has been added, in-phase, to both fields.  The initial energy $E_0$ is indicated by color.}
  \label{fig:epsvsfrac}
\end{figure}

Figure~\ref{fig:epsvsfrac} shows the results of a Monte Carlo sampling for $\ICsucc$, analogous to that in Figure~\ref{fig:failepsvsfrac}, with one initially excited mode added in-phase to both $\psi$ and $\phi$.  We again use a logarithmic prior, but with the ranges: $-2.0 < \log_{10} f < 0.0 $ and $-3.0 < \log_{10} A < 0.0 $.  Adding perturbations to $\ICsucc$ with initially small gradient energies does not cause the spatially averaged trajectories in field-space to deviate significantly from the homogeneous solution, as seen in Fig.~\ref{fig:trajs}, and these configurations successfully inflate.  However, if we add more significant inhomogeneity with amplitude $A \gtrsim 0.1$, then any transitory inflation is typically disrupted before $N_e \sim 5$.  We do not see any indication of a fractal structure in the distribution of successful configuration in $(A,f)$-space for this initial condition.

\subsection{Multiple excited modes}
\label{ssect:multiplemodes}

We now turn to the more general case, with multiple excited modes in both fields $\phi$ and $\psi$:
\begin{equation}
  \phi_i(0,x) = \bar \phi_{i,0}\left[1 + \sum_{j=1}^\mathcal{N} A_{ij} \sin \left( \frac{2 \pi n_{ij} x}{L} + \alpha_{ij} \right) \right] \, ,
  \label{eqn:multipleinhom}
\end{equation}
where $A_{ij}$ is the real amplitude and $\alpha_{ij}$ is the phase of the $j\th$ mode for the $i\th$ field. The box size $L$ is set to the wavelength of the largest mode of interest by fixing $f$ via Eq.~\eqref{eqn:frac}.  The integer $n_{ij}$ gives the ratio of the $i\th$ field's $j\th$ mode's wavelength, relative to the largest mode.  Each field is assumed to have  $\mathcal N$ excited modes.

\begin{figure}
  \centering
  \includegraphics{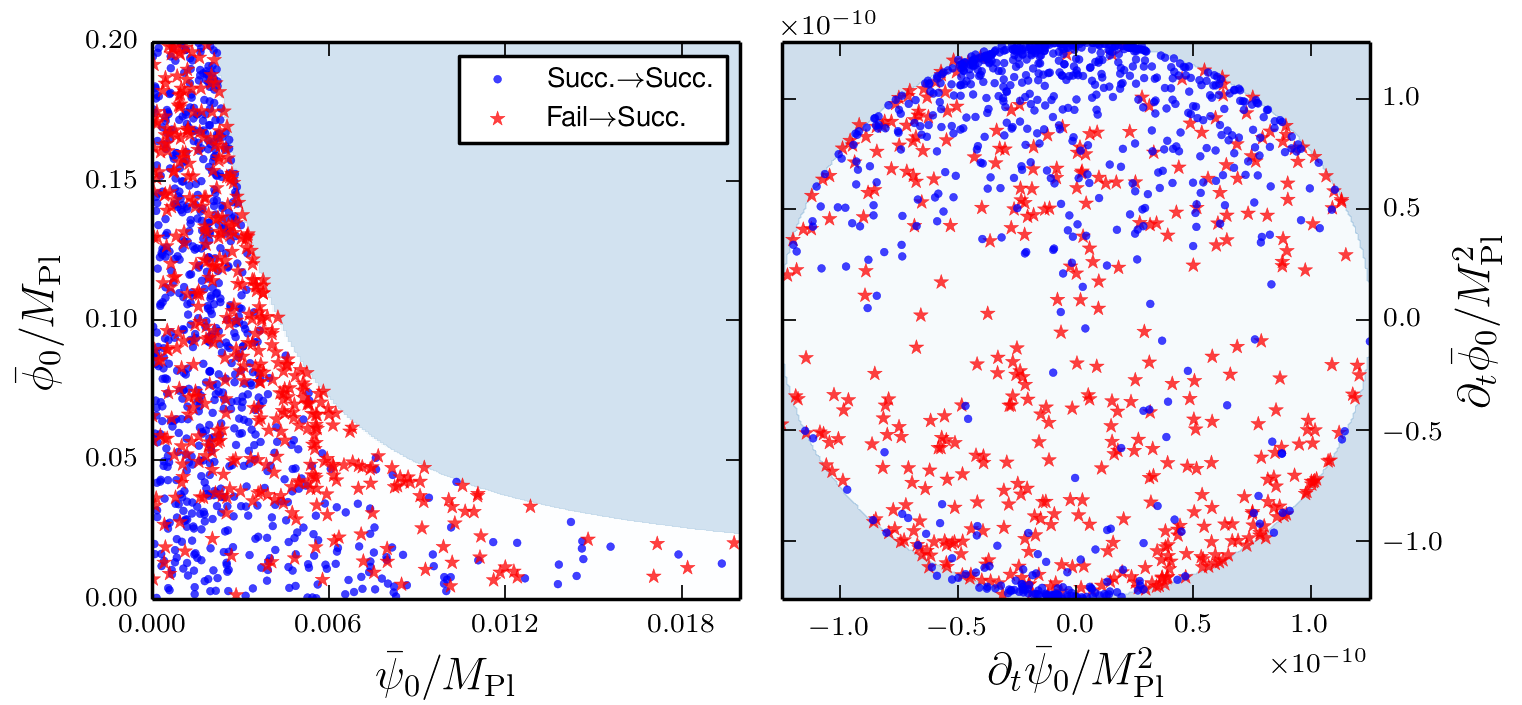}
  \caption{Distribution of successful ICs in (\emph{left}) field-space and (\emph{right}) velocity-space that are (\emph{blue dots}) successful and (\emph{red stars}) unsuccessful at giving $N_e >55$ in the homogeneous limit.  The homogeneous ICs are chosen with $E_0=10^{-5} \, \mpl$.  The ICs have two initially excited modes of different scales.  (\emph{Background})  The gray region has not been sampled since the initial energy density would have exceeded $E_0^4$.  For the velocities, the background has also been offset by $\Lambda^4$.
  }
  \label{fig:fields_multi}
\end{figure}

\begin{figure}
  \centering
  \includegraphics{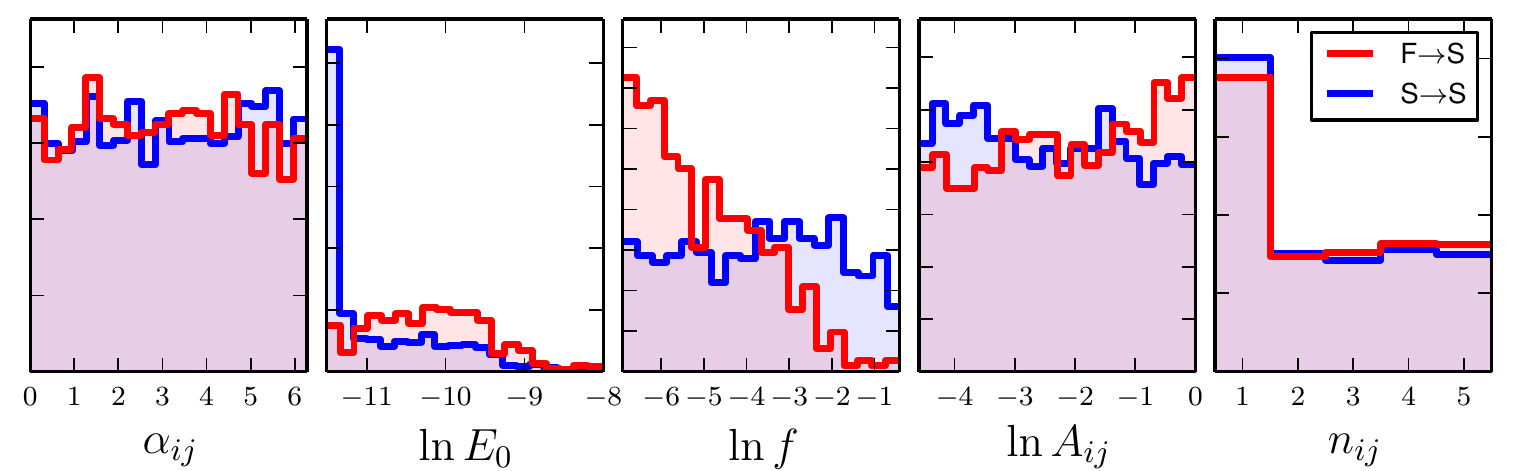}
  \caption{Histograms of marginalized inhomogeneity parameters from Eq.~\eqref{eqn:multipleinhom} for successfully inflating ICs; the data for both fields have been plotted together.  Two modes have been initially excited in each field and we plot ICs that both (\emph{blue}) succeed and (\emph{red}) fail to inflate in the homogeneous limit.  The large amplitude bin at $n=1$ (\emph{far-right}) results from setting the largest mode to $n=1$ and is therefore systematic.
  }
  \label{fig:hist_multi}
\end{figure}

Figs~\ref{fig:fields_multi}~and~\ref{fig:hist_multi}  plot the results of a Monte Carlo analysis with $\mathcal N=2$  excited modes.  We choose the background values $(\bar \psi, \bar \phi, \partial_t {\bar \psi}, \partial_t{ \bar \phi})$ so that each configuration has initial energy $E_0=10^{-5} \, \mpl$ using the \emph{iso-$E$} measure of Refs~\cite{Easther:2013bga,Easther:2013rva}.  In general, multifield models make predictions for observables that are largely independent of the prior probability distribution of the ICs \cite{Frazer:2013zoa,Easther:2013bga,Easther:2013rva}, so this choice should not have a large effect on the results.  We then add initial inhomogeneity, which marginally increases the overall energy. We draw the amplitudes from a logarithmic prior, $10^{-2} < A_{ij} < 1$, and the largest mode with a logarithmic prior in the range $10^{-3} < f < 1$.  The wavelength of the second mode is drawn uniformly from the range $1< n_{ij} < 5$ with uniform random phase $0 < \alpha_{ij} < 2 \pi$.  These plots can be compared to Fig.~5 in Ref.~\cite{Easther:2013bga}, which presents histograms of successful ICs in the homogeneous limit.

Figure~\ref{fig:fields_multi} shows successful initial configurations projected onto both the homogeneous field space $\{ \bar{ \psi}_0, \bar{ \phi}_0\}$ and the homogeneous velocity space $\{ \partial_t \bar{ \psi}_0, \partial_t \bar{ \phi}_0\}$.  When including initial inhomogeneity, successful configurations are approximately uniformly distributed for $\bar \psi_0 \lesssim 0.005 \, \mpl$.  There is some minor difference between the location of ICs that succeed and those that fail, which is primarily due to the the fact that ICs closer to $\psi \approx 0$ tend to be more likely to inflate as they are closer to the inflationary valley.  In velocity space, there is a tendency for successful ICs to have $\partial_t \bar{ \phi}_0 \sim 10^{-10} \, M_\mathrm{Pl}^2$ and $\partial_t \bar{ \psi}_0 \approx 0$, since having a large velocity in $\psi$ causes the trajectory to evolve away from the inflationary valley at $\vev \psi \approx 0$.  This behavior again matches the homogeneous limit \cite{Easther:2013bga}, as the blue points in Fig.~\ref{fig:fields_multi} cluster in this range.  Again, many configurations that fail in the homogeneous limit succeed when multiple modes are initially excited.

Fig.~\ref{fig:hist_multi} displays normalized histograms of the inhomogeneity parameters $A_{ij}$, $n_{ij}$, and $\alpha_{ij}$.  Because we have not been careful to sample unique solutions to Eq.~\eqref{eqn:kgeqn}  \cite{Corichi:2010zp,Easther:2013bga,Corichi:2013kua} and because we have used hybrid inflation~\eqref{eqn:hybridv} only as a toy model, we do not give a detailed analysis of the inhomogeneity parameters in each field, but rather plot the values together.  The histograms do not show any dependence on the phases $\alpha_{ij}$, indicating that the results in Section~\ref{ssect:singlemode}, which use in-phase inhomogeneities, are robust.  There is a large peak in the mode number $n_{ij}$ at the largest wavelength; however, we have forced at least one mode to have $n_{ij} = 1$ and set the other modes with wavelengths with integer multiples of this largest mode.  Consequently, the spike at $n_{ij}=1$ results from systematics only and there is no strong dependence on mode number for successful ICs that are either successful or unsuccessful in the homogeneous limit.  Because we fix the background energy scale, the histogram for $E_0$ directly measures the initial gradient energy, which we allow to be up to 35 times the initial homogeneous energy.  

Initial configurations that are successful in the homogeneous limit tend to also be successful with initial inhomogeneity, provided the initial gradient energy is relatively small.  Again, this can be understood in terms of Fig.~\ref{fig:trajs}, as the trajectories with small initial gradients are indistinguishable from the homogeneous trajectory.  However, many initial configurations that are successful in the homogeneous limit remain successful with a large initial gradient, although the fraction decreases with increasing initial gradient energy.

Points for which inflation fails in the homogeneous limit have a strong dependence on the size of the largest initially excited mode $f$ and a weak dependence on mode amplitude $A_{ij}$, favoring higher initial gradient energy.  The number of successful configurations decreases when the initial energy exceeds $E_0 \gtrsim 7.5\e{-5} \mpl$.  Given that our lattice simulation technique will be unreliable with large gradient energies, these cases need to be investigated within the full Einstein equations.

\section{Conclusion}
\label{sect:conclusion}

We have demonstrated that adding subhorizon inhomogeneity can significantly modify the pre-inflationary dynamics of hybrid inflation, yielding counterintuitive results.  While the solutions to the Klein-Gordon equation are generally stable to the addition of small initial inhomogeneity, if the inhomogeneity has a larger amplitude, then the evolution changes dramatically.  Unsurprisingly, large primordial inhomogeneities may prevent the onset of inflation in models which do inflate in the homogeneous limit. However, this is not a certainty: models which do not inflate in the homogeneous limit can successfully inflate when inhomogeneity is added, since the spatially averaged trajectories traverse more of phase space and have a greater chance of inflating via a different path in phase-space.  We have presented an ensemble Monte Carlo analysis with varying types of initial inhomogeneity to demonstrate that this behavior is generic for hybrid inflation and should be true for any multifield inflation model that has chaotic behavior.

We have analysed the toy hybrid inflation model defined by Eq.~\eqref{eqn:hybridv}, but  argue that this behavior should be common for multifield inflation models that have unstable fixed points or saddle points in the potential.  This model does not yield the correct perturbation spectrum, but this is not a significant issue given that our focus here is the onset of inflation.

The solutions to the inhomogeneous Klein-Gordon equation~\eqref{eqn:kgeqn} are still qualitatively chaotic when the spatially averaged field trajectories are plotted.   This paper thus provides the first confirmation that the chaotic dynamics extend from the ordinary differential equations of the homogeneous problem to the partial differential equations of the inhomogeneous universe.  The chaotic dynamics result from the interplay between the fixed points and the saddle point at the critical value for $\phi$.  Consequently, our qualitative conclusions should extend to other models with these features.

It was argued in Refs.~\cite{Vachaspati:1998dy,Trodden:1999wc} that the conditions necessary to start inflation must extend over a super-Hubble region.  Here, we have demonstrated that we do not need to necessarily require homogeneity and that significant sub-horizon perturbations may not prevent inflation from beginning, despite the conventional wisdom.  This work clearly has a number of possible extensions, both to wider classes of models and also to include the effects of non-zero curvature, which is generically expected in pre-inflationary scenarios \cite{Berera:2000xz,Linde:2004nz}.  Furthermore, including local gravitational backreaction by solving the full Einstein field equations would extend this analysis to configurations with large or asymmetric gradient energies.

\acknowledgments

We thank Grigor Aslanyan for comments on the manuscript and Gary Felder and Igor Tkachev for making \textsc{LatticeEasy} [\url{www.cita.utoronto.ca/~felder/latticeeasy/}] freely available.  We acknowledge the contribution of the NeSI high-performance computing facilities and the staff at the Centre for eResearch at the University of Auckland. New Zealand's national facilities are provided by the New Zealand eScience Infrastructure (NeSI) and funded jointly by NeSI's collaborator institutions and through the Ministry of Business, Innovation \& Employment's Research Infrastructure programme [{\url{http://www.nesi.org.nz}}].  JR acknowledges financial support from Spanish MEC and FEDER (EC) under grant FPA2011-23596.

\bibliographystyle{JHEP}

\bibliography{references}

\end{document}